\documentclass{iau} 
\usepackage{graphicx}

\newcommand{\vsini}{$v_{eq} \sin i$}
\newcommand{\vinf}{$v_\infty$}
\newcommand{\iue}{\textit{IUE}}

\newcommand{\xmm}{\textit{XMM}}
\newcommand{\ha}{H$\alpha$}
\newcommand{\mdot}{{$\dot{M}$}}

\newcommand{\hst}{{\textit{HST}}}

\newcommand{\nv}{N~{\sc v}}

\newcommand{\niv}{N~{\sc iv}}

\newcommand{\siiv}{Si~{\sc iv}}
\newcommand{\siiii}{Si~{\sc iii}}

\newcommand{\kms}{km s$^{-1}$}
\newcommand{\taurad}{{$\tau_{rad}$}}
\newcommand{\fratio}{$f$-ratio}
\newcommand{\apj}{\textit{Ap.J.}}
\newcommand{\apjs}{\textit{Ap.J. Supp.}}
\newcommand{\apjl}{\textit{Ap.J. Letters}}
\newcommand{\aap}{\textit{A\&A}}
\newcommand{\mnras}{\textit{M.N.R.A.S.}}

\title[Wind line variability] 
{Wind line variability and intrinsic errors in observational mass loss rates}

\author[Derck Massa \& Raman Prinja]   
{Derck Massa$^1$
 \and Raman Prinja$^2$}

\affiliation{$^1$Space Science Institute, \\ 4750 Walnut Street 
Suite 205 Boulder, Colorado 80301, USA
\\ email: {\tt dmassa@spacescience.org} \\[\affilskip]
$^2$Department of Physics \& Astronomy,  
University College London,\\ Gower Street, London WC1E 6BT, UK
\\email: {\tt rkp@star.ucl.ac.uk}}

\pubyear{2015}
\volume{xxx}  
\jname{Title of your IAU Symposium}
\editors{A.C. Editor, B.D. Editor \& C.E. Editor, eds.}
\begin{document}

\maketitle

\begin{abstract}
UV wind line variability in OB stars appears to be universal.  We review 
the evidence that the variability is due to large, dense, optically thick 
structures rooted in or near the photosphere.  Using repeated observations 
and a simple model we translate observed profile variations into optical 
depth variations and, consequently, variations in measured mass loss rates.  
Although global rates may be stable, measured rates vary.  Consequently, 
profile variations infer how mass loss rates determined from UV wind lines 
vary.  These variations quantify the intrinsic error inherent in any mass 
loss rate derived from a single observation.  These derived rates can 
differ by factors of 3 or more.  Our results also imply that rates from 
non-simultaneous observations (such as UV and ground based data) need not 
agree.  Finally, we use our results to examine the nature of the structures 
responsible for the variability.

\end{abstract}

\firstsection 
\section{Introduction}

Wind line variability, as seen in \ha, has been known for many years 
\cite[(e.g., Ebbets 1982)]{ebbets82}.  With the advent of \iue, it became 
clear that well developed but unsaturated UV wind lines in OB stars were 
also highly variable \cite[(e.g., Prinja \& Howarth 1986)]{prinja86}.  
It is important to study this wind line variability for the following 
reasons: 1) It appears to be a universal property of radiatively driven 
winds. 2) Currently there are four ways to measure mass loss rates (\mdot): wind 
lines, IR/Radio excesses, X-rays and bow shocks.  They do not all agree, 
due to structures (clumping) in the winds (e.g., \cite[Fullerton \etal\ 
2006]{fullerton06}; \cite[Massa \etal\ 2017]{massa17}; Kobulnicky \etal\ 
2019 AJ, 158, 73) and wind line variability \textit{presents an opportunity 
to characterize these wind structures and provide feedback for modelers.}.  
3) Wind line variability \textit{determines the intrinsic accuracy of a 
single measurement of \mdot.}  

In the following, we discuss what is known about the nature and origin of 
the variability.  We then introduce a simple model, use it to analyze the 
variability in a few times series and then summarize our results.  

\section{Observations of wind line variability} 
\begin{figure}
\begin{center}
\includegraphics[width=0.7\linewidth]{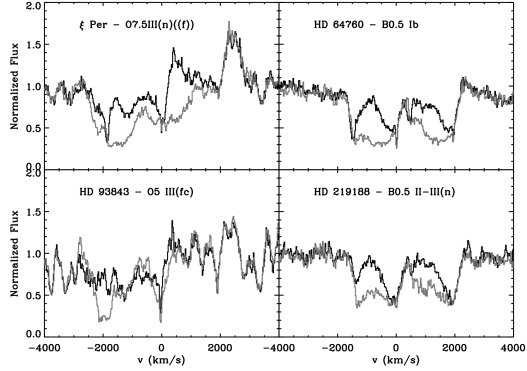}
\end{center}
\vspace{-2.2in}
\caption{
 Examples of \siiv $\lambda\lambda 1400$\ variability in four stars with a 
 range in spectral types.}
\label{fig:var}
\end{figure}

\begin{figure}[h]
\begin{center}
\includegraphics[width=0.8\linewidth, angle=180]{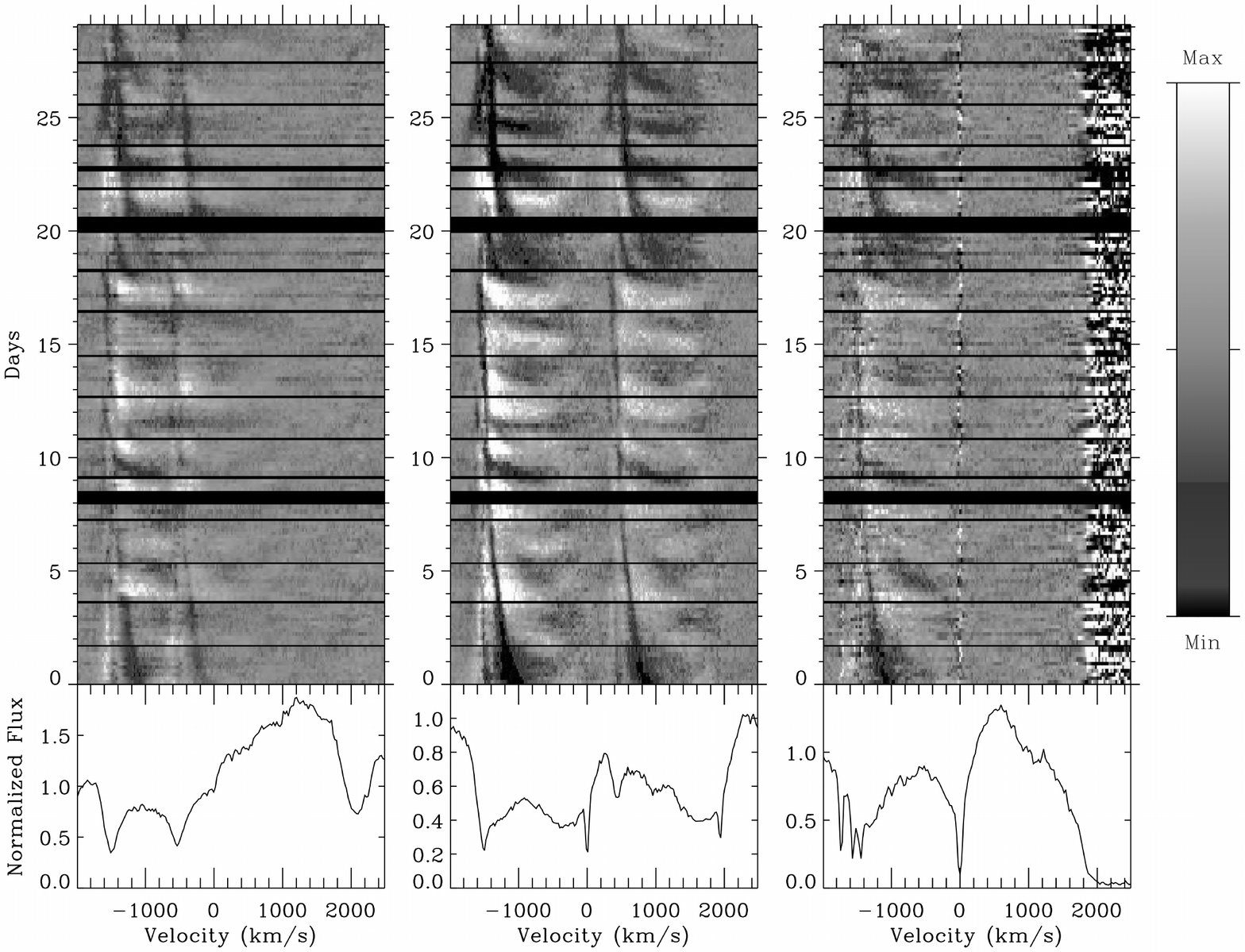}
\end{center}\vspace{-0.25in}
\caption{Dynamic spectra of (left to right) \nv, \siiv\ and \siiii\ for 
HD 64760 (B0.5 Ib, $v \sin i = 216$\ \kms. 
\cite{massa95}.  
}\label{fig:mega}\end{figure}

In this section we describe what has been learned from the morphology of 
wind line variability.  Figure~\ref{fig:var} shows examples of UV wind line 
variability in 4 stars with a range in spectral types, but all of which 
have well developed, but unsaturated \siiv\ $\lambda 1400$ resonance 
doublets.  Figure~\ref{fig:mega} is the dynamic spectrum (a gray scale of 
all the spectra in a series divided by the mean) of a time series of the 
B0.5~Ib star, HD 64760, described by \cite{massa95}.  Such data sets have 
lead to the following conclusions.

\noindent {\bf Wind variability is universal:} Wind line variability is 
seen in repeated observations of just about every OB star with a well 
developed but unsaturated wind line \cite[(e.q. Prinja \& Howarth 1986)]
{prinja86}.  Further, it has also been observed in LMC and SMC OB stars 
(\cite[Massa \etal\ 2000]{massa00}) and the central stars of planetary 
nebulae (\cite[Prinja \etal\ 2012]{prinja12}).

\noindent {\bf Spiral structures cause the variability:}  Time series 
of UV wind lines revealed bow shaped patterns in the dynamic spectra 
OB Stars (see Figure~\ref{fig:mega}).  Analysis of these patterns lead 
\cite{cranmer96} to interpret them in terms of co-rotating interaction 
regions (CIRs).  However, any phenomenon that produces a spiral structure 
in the wind will do. To see the classic bow shape, one must view the star 
with $\sin i \simeq 1$\ and the origin of the spot must be near the 
equator.  Consequently, it is not surprising that the pattern is not seen in 
some stars with lower \vsini\ s, although regularly repeating patterns are 
(\cite[Prinja \etal\ 2002]{prinja02}).

\noindent {\bf The spiral structures are optically thick:} \cite{prinja10} 
applied a simple model to show that the \siiv\ $\lambda 1400$ doublet wind 
line profiles of a large sample of B supergiants are best fit using doublet 
ratios that are between unity and the actual ratio of the oscillator 
strengths of the doublet, $f_B/f_R \simeq 2$.  This indicates that the line 
of sight to the stellar disk is partially covered by very optically thick 
structures, and by optically thin material.  The result is a doublet 
absorption that appears unsaturated (because it does not go to zero) but 
whose relative strength is close to unity (since most of the absorption is 
from the strongly saturated optically thick component).   

\begin{figure}[h]
\begin{center}
\includegraphics[width=0.8\linewidth, angle=180]{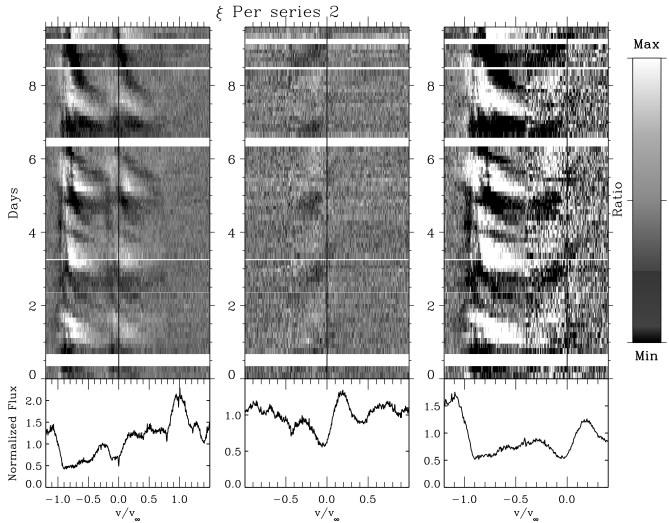}
\end{center}\vspace{-0.25in}
\caption{{Dynamic spectra of \siiv, \niv, and the two spliced together for 
  $\xi$ Per (\vinf = 2450 \kms, \cite[(Massa \& Prinja 2015)]{massa15}.  
  Shows that structures are tied to the base of the wind where the radiation 
  field is intense.} }
  \label{fig:dynamic}
\end{figure}

\noindent {\bf The spiral structures reach the stellar surface and are 
large:} The fact that the variability appears to extend to $v \simeq 0$, 
does not necessarily mean that the physical structures extend to $r \simeq 
R_\star$, since this could simply reflect a non-monotonic velocity law. 
However, \cite{massa15} showed that the absorption of the excited state 
line \niv~1718 blends smoothly with the high velocity absorption in 
resonance lines (see, Fig~\ref{fig:dynamic}).  Since \niv~1718 cannot exist 
without the strong photospheric radiation field, this analysis shows that 
the features must originate near $r = R_\star$.  Furthermore, since the 
features cause detectable absorption near $R_\star$, they must occult a 
significant portion of the stellar disk.  

\noindent {\bf The spiral structures are denser than ambient wind:} A large 
change in the optical depth can indicate either a change in density, a 
flattening of the velocity law, or a change in the ionization (see, \S 
\ref{sec:sei}).  It is, therefore, important to examine observational 
diagnostics that respond only to density.  X-ray fluxes are one such 
diagnostic.  It has been known for some time that the X-ray fluxes in OB 
stars are variable \cite[(see Oskinova \etal\ 2001, ]{oskinova01} 
\cite[Naz\'e \etal\ 2013, ] {naze13}  \cite[Naz\'e \etal\ 2018)]{naze18}.  
Most recently, \cite{massa19} obtained contemporaneous \xmm\ X-ray and \hst\ 
STIS UV spectra of the O7.5 III(n)((f)) star $\xi$~Per.  They were able to 
show that the X-rays are modulated by the spiral patterns, implying that 
they are significantly denser than the rest of the wind (the exact amount 
is model dependent).


\vfill

\vfill

\section{Modeling wind line variability}\label{sec:sei}

In order to translate flux variations into the physical parameters needed to 
estimate the intrinsic error in a single \mdot\ measurement, a model is 
required.  We use the SEI Sobolev model as formulated by \cite{lamers87} and 
modified by \cite{massa03}. 

\noindent The calculation of an SEI profile requires the following 
parameters: 
\begin{enumerate}
\item[1)]  A value for {\vinf} 
(determined by a grid search), 
\item[2)] {A velocity law:} typically a 
{$\beta$}-law, of the form {$v = v_\infty (1 - a/x)^\beta$}, where 
{$x = r/R_\star$}. 
\item[3)] The radial (Sobolev) optical depth of the wind ($w = v/$\vinf): 
$$
  {
  \tau_{rad}(w) = Const \frac{\dot{M}}{R_\star v_{\infty}^2} 
                q_i(w) \left(x^2 w \frac{dw}{dx}\right)^{-1} 
                }
$$ 
where $Const$\ contains atomic parameters and $q_i$ is the ionization fraction.  
$\tau_{rad}(w)$ is modeled by 20 velocity bins adjusted to obtain 
the best fitting profile.  This approach can be viewed as an inversion of 
the profile to obtain $\tau_{rad}(w)$.  \textit{Note that $\tau_{rad}$\ 
variations are proportional to derived $\dot{M}q$\ variations.}

\end{enumerate}

In addition to the usual parameters, we also allow the ratio of the optical 
depths of the doublets, \fratio\ $ \equiv f_B/f_R$, to vary.  This is a 
well known means to mimic the effect of optically thick structures 
partially covering the stellar disk \cite[(Prinja \& Massa 2015)]{prinja15}.  
It also provides an additional diagnostic of how the portion of the wind 
structures in front of the star varies with time.  We also note that it has 
little affect on \taurad.  Its main effect is to improve the SEI fit to the 
red component of the wind line.  All of these parameters are determined by a 
non-linear least squares fit to the observed profile.

\section{Results}

\begin{figure}
\begin{center}
\includegraphics[width=0.4\linewidth]{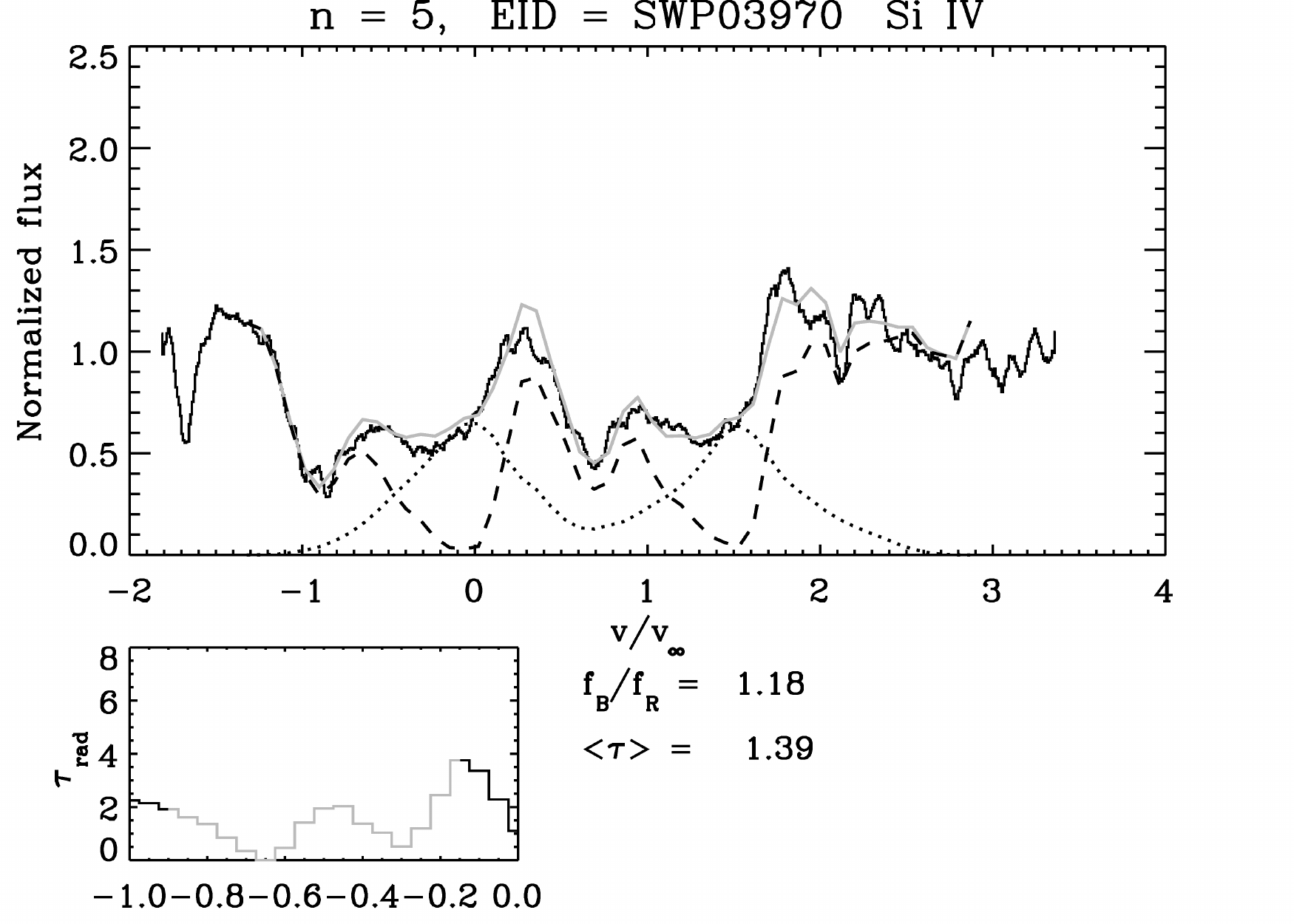}
\includegraphics[width=0.4\linewidth]{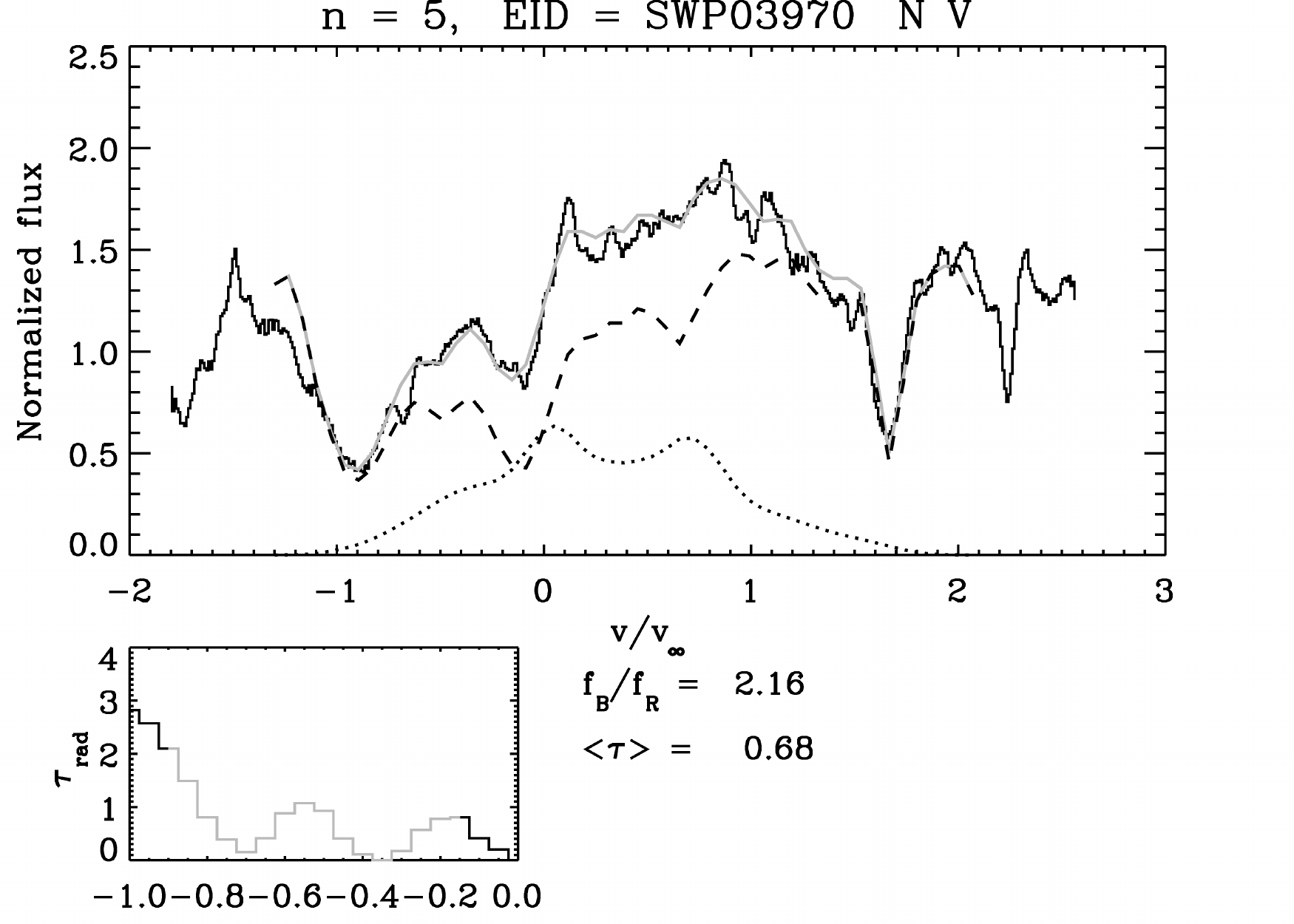} 
\end{center}
\caption{Fits to a \siiv\ and \nv\ profiles in {$\rho$}\ Leo with derived 
  \taurad s below.  Observed = solid black, fit = gray, absorbed 
  photospheric profile = dashed and emission = dotted.  }
\label{fig:fits}
\end{figure}

Figure~\ref{fig:fits} shows the quality of the fits that can be achieved.  
It shows non-linear least squares fits to the \siiv\ and \nv\ profiles in 
{$\rho$}\ Leo.  The model used \vinf = 1150 \kms, and a TLUSTY \cite[(Lanz \& 
Hubeny 2003)]{lanz03} model with solar abundance and $(T_{eff}, \log g) = 
(25kK, 2.5)$\ for the photospheric spectrum.  The derived \taurad s are 
shown below each spectrum, and the portion used to determine $<\tau>$, $0.2 
\leq v/v_\infty < 0.9$, is shaded gray.  The best fit \fratio\ and $<\tau>$\ 
values are also listed.  Notice that the fit with the smaller $< \tau >$\ 
has the larger \fratio. 

As part of an ongoing program to analyze all repeated \iue\ observations of 
normal stars with well developed but unsaturated lines, we have begun with 
an analysis of a few time series.  Figure~\ref{fig:relations} summarizes the 
results for two time series, one consisting of 70 spectra of $\xi$~Per and 
one for 146 spectra of HD~64760.  The left hand panels show the mean 
\taurad\ s and \fratio s versus time, while the right hand panels show the 
two quantities plotted against each other.    

\section{Analysis}

The previous results can be used to address two distinct issues: the 
intrinsic error present in any \mdot\ derived from a single observation, and 
the nature of the structures responsible for the variations.  

\begin{figure}
\begin{center}
\includegraphics[width=0.41\linewidth]{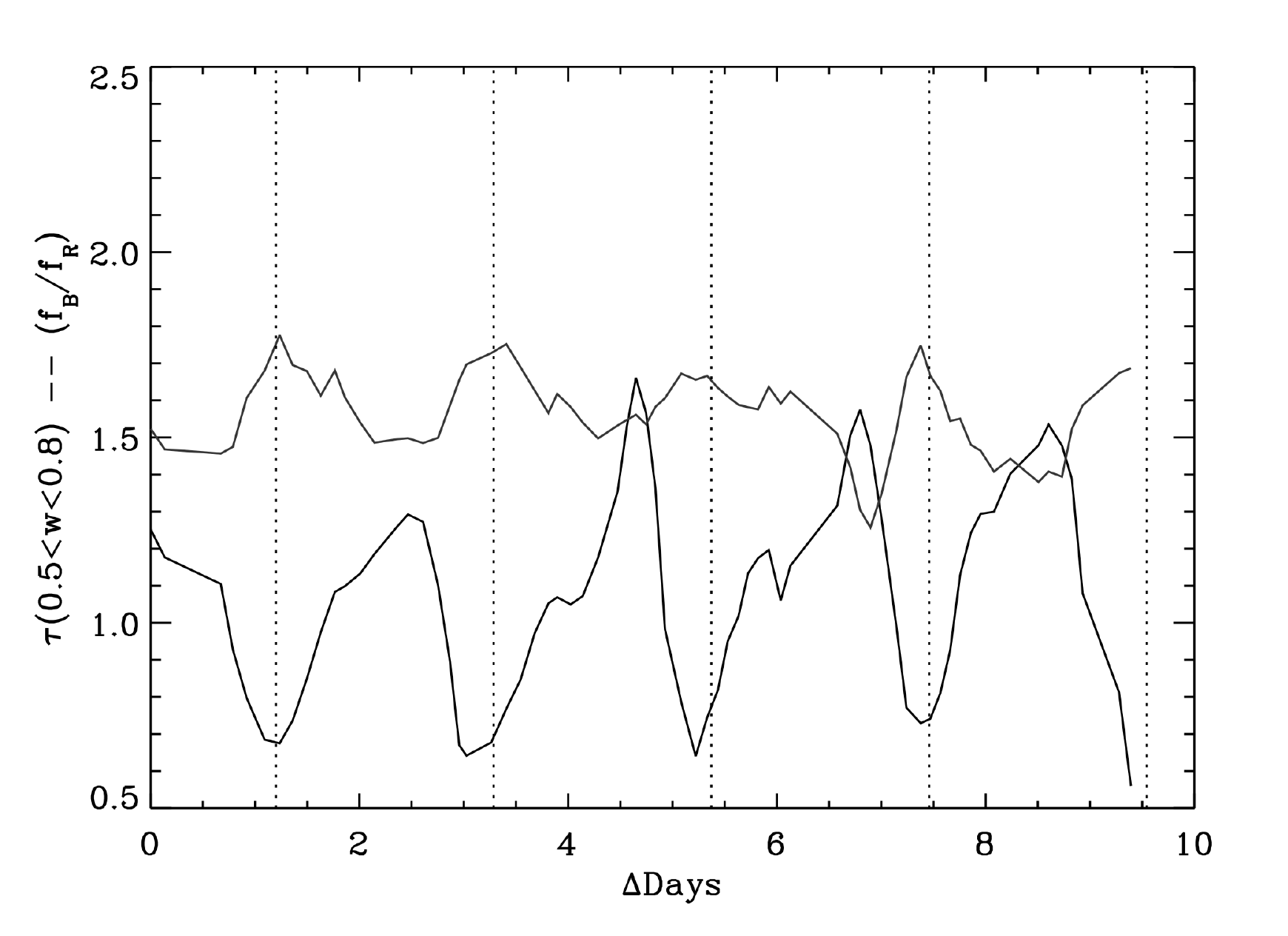}
\includegraphics[width=0.45\linewidth]{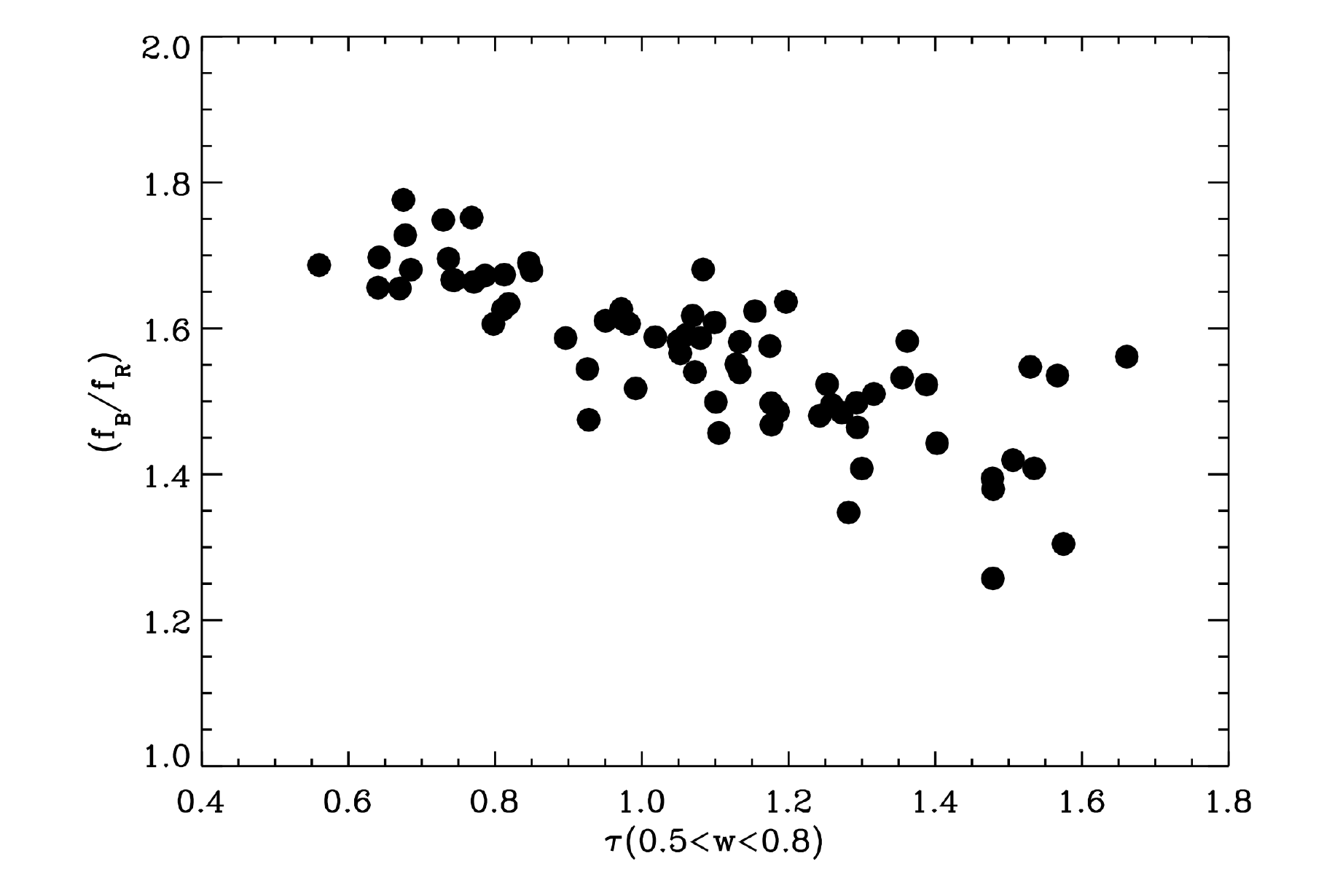} \vspace{-2in}
\includegraphics[width=0.45\linewidth]{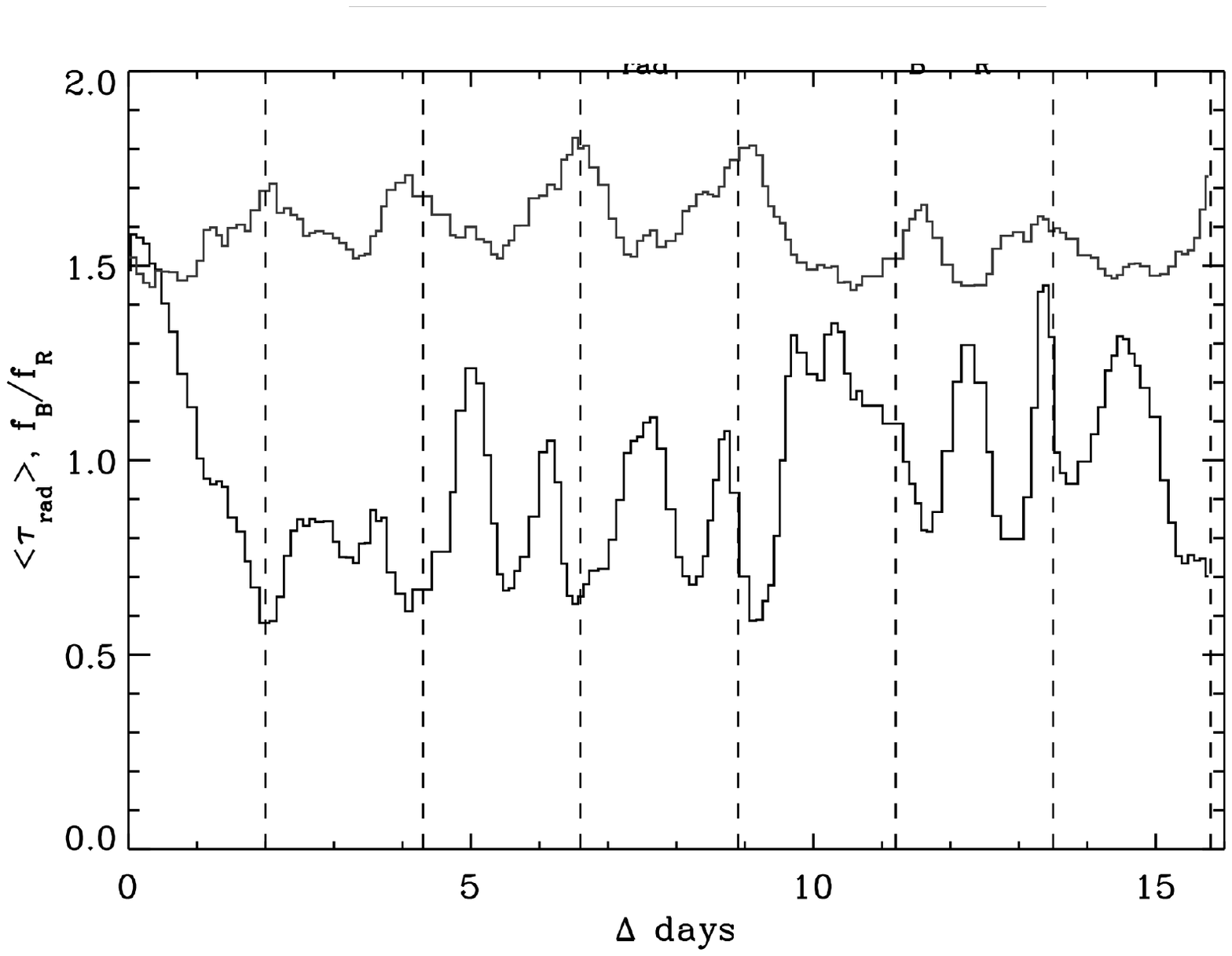}
\includegraphics[width=0.45\linewidth]{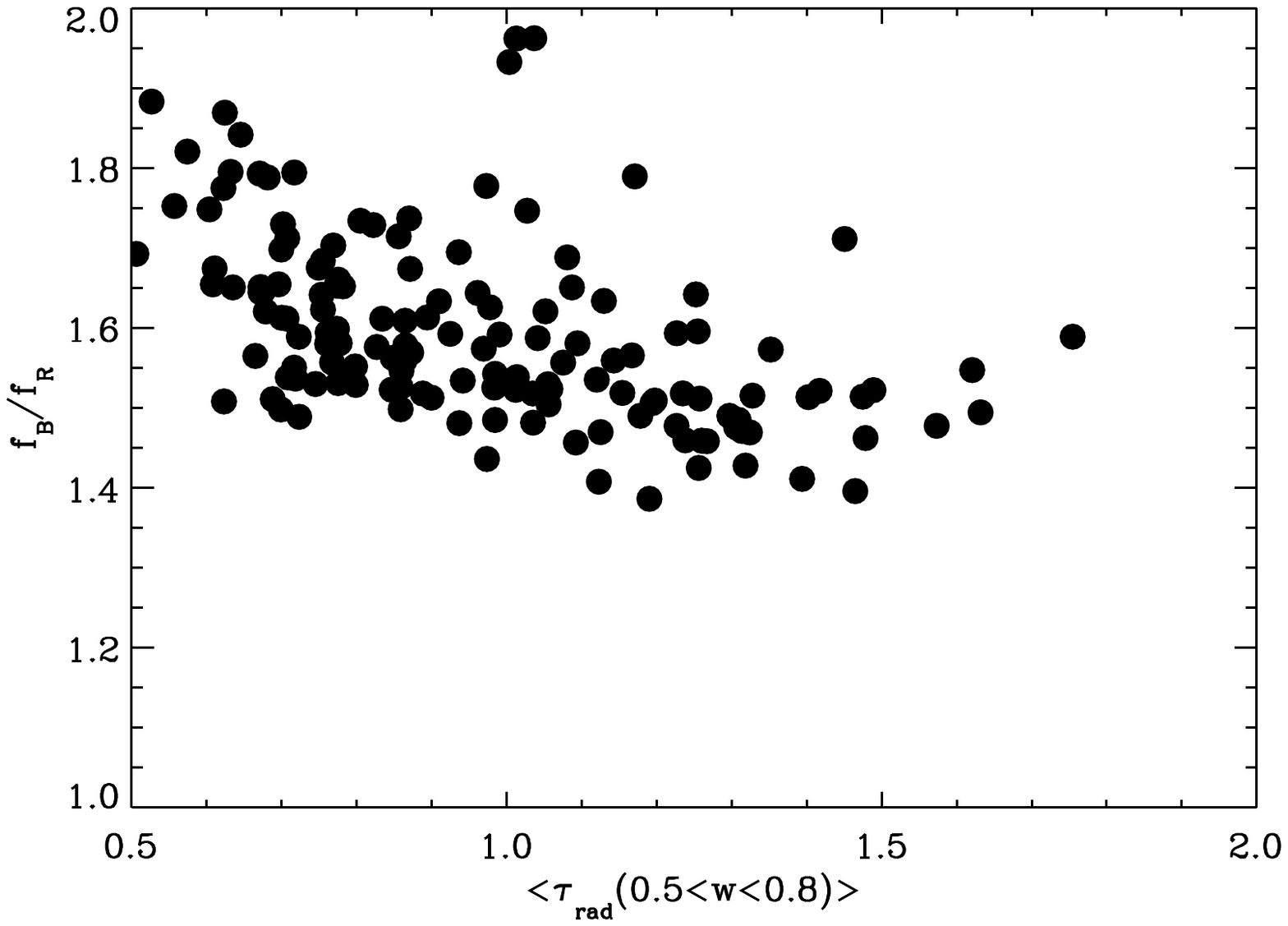} 
\end{center}
\vspace{.5in}
\caption{
  {\bf Top row:} 70 \siiv\ spectra of $\xi$\ Per.  Left: Variations of 
  \taurad\ averaged over $0.5 \leq v/v_\infty \leq 0.8$\ (lower curve) and 
  \fratio s (upper curve) versus days from first spectrum.  Dotted vertical 
  lines are the 2.086 day period determined by \cite{dejong01}.  Right: The 
  same two quantities plotted against each other.  {\bf Bottom row:} 146 
  \siiv\ spectra of HD 64760. Left: Mean \taurad\ s (lower curve) and 
  \fratio s (upper curve) versus days from the first spectrum.  Dotted 
  vertical lines are a 2.30 Day period. Right: The same two quantities 
  plotted against each other. }
\label{fig:relations}
\end{figure}

\noindent \textbf{The intrinsic error:} Recall that \taurad\ is directly 
proportional to \mdot, so variations in \taurad\ are a surrogate for 
variations in \mdot.  Examination of the right panels of 
Figure~\ref{fig:relations} shows that the ranges of $<\tau_{rad}>$ can be a 
factor of 3 or more.  The means and variances of $<\tau_{rad}>$ for the 
$\xi$~Per series are  = 1.09 and 0.27 and for the HD~64760 series they are 
1.05 and 0.26. Both cases suggest the best accuracy one can expect is about 
$\pm 25$\%.  However, this may be a lower limit since the distributions are 
highly non-Gaussian and the fact that the observations were clustered in 
time may introduce an additional bias.  Nevertheless, a reasonable first 
estimate is $\sigma(\dot{M})/\dot{M} \simeq 1.25$, although this is 
probably a lower limit.  

\noindent \textbf{Constraints on the structures:} It is interesting that, 
for the two series analyzed, the \fratio\ varies periodically, as distinctly 
as the $<\tau_{rad}>$ or even more so.  Further, the \fratio\ decreases as 
$<\tau_{rad}>$ increases, suggesting that the change in the apparent optical 
depth is actually due to a larger fraction of the stellar surface being 
covered by optically thick material. These results are intriguing, but we 
emphasize that they are very preliminary.

\section{Summary}

We reviewed the evidence that wind line variability is universal and 
probably due to large, spatially coherent spiral structures.  These 
structures extend to the base of the wind and cover much of the stellar 
disk. Further, they are denser than their surroundings, and because they 
are optically thick, {\em their geometry matters}.  We then showed that the 
variability limits the intrinsic accuracy of a single \mdot\ measurement 
derived from a resonance line to roughly 25\%, and that individual 
measurements can vary by as much as a factor of 3.  This variability also 
compromises the consistency of non-simultaneous measurements.  Finally, we 
discussed how our results can be used to constrain the nature of the 
structures responsible for the variability.



\begin{discussion}



\end{discussion}


\begin{thebibliography}{}

\bibitem[Cranmer \& Owocki(1996)]{cranmer96} Cranmer, S.R. \& Owocki, S.P.\ 
  1996, \apj, 462, 469 

\bibitem[Ebbets(1982)]{ebbets82} Ebbets, D.\ 1982, \apjs, 48, 399  

\bibitem[de Jong \etal\ (2001)]{dejong01} de Jong, J.A., Henrichs, H.F., 
Kaper, L., \etal\ 2001, \aap, 368, 601 

\bibitem[Fullerton \etal\ (2006)]{fullerton06} Fullerton, A.W., Massa, D.L., 
\& Prinja, R.K.\ 2006, \apj, 637, 1025 

\bibitem[Lamers \etal\ (1987)]{lamers87} Lamers, H.J.G.L.M., Cerruti-Sola, 
  M., \& Perinotto, M.\ 1987, \apj, 314, 726 

\bibitem[Lanz \& Hubeny (2003)]{lanz03} Lanz, T. \& Hubeny, I.\ 2003, \apjs, 
146, 417 

\bibitem[Massa \etal\ (1995)]{massa95} Massa, D., Fullerton, A.W., Nichols, 
  J.S., et al.\ 1995, \apjl, 452, L53 

\bibitem[Massa \etal\ (2000)]{massa00} Massa, D., Fullerton, A.W., 
  Hutchings, J.B., \etal\ 2000, \textit{ApJL}, 538, L47 

\bibitem[Massa \& Prinja (2015)]{massa15} Massa, D. \& Prinja, R.K.\ 2015, 
\textit{ApJ}, 809, 12 

\bibitem[Massa \etal\ (2019)]{massa19} {Massa, D., Oskinova, L., Prinja, R. 
\& Ignace, R.} 2019, \apj, 873, 81

\bibitem[Massa \etal\ (2003)]{massa03} Massa, D., Fullerton, A.W., 
  Sonneborn, G., et al.\ 2003, \apj, 586, 996 

\bibitem[Naz{\'e} \etal\ (2013)]{naze13} Naz{\'e}, Y., Oskinova, 
  L.M., \& Gosset, E.\ 2013, \apj, 763, 143 

\bibitem[Naze{\'e} \etal\ (2018)]{naze18} Naz{\'e}, Y., Ramiaramanantsoa, 
  T., Stevens, I.R., \etal\ 2018, \aap, 609, A81  

\bibitem[Oskinova \etal\ (2001)]{oskinova01} Oskinova, L.M., Clarke, D., 
  \& Pollock, A.M.T.\ 2001, \aap, 378, L21 

\bibitem[Fullerton \etal\ (1997)]{fullerton97} Fullerton, A.W., Massa, D.L., 
  Prinja, R.K., \etal\ 1997, \aap, 327, 699  

\bibitem[Prinja \& Howarth (1986)]{prinja86} Prinja, R.K. \& Howarth, I.D.\ 
1986, \apjs, 61, 357 

\bibitem[Prinja \etal\ (2012)]{prinja12} Prinja, R.K., Massa, D.L., 
  Urbaneja, M.A., \etal\ 2012, \mnras, 422, 3142 

\bibitem[Prinja \etal\ (2002)]{prinja02} Prinja, R.K., Massa, D., \& 
  Fullerton, A.W.\ 2002, \aap, 388, 587 

\bibitem[Prinja \& Massa (2010)]{prinja10} Prinja, R.K. \& Massa, D.L.\ 
2010, \aap, 521, L55 

\end{thebibliography}
\end{document}